\documentclass{PoS}
\usepackage{url}
\usepackage[sort]{natbib}
\bibpunct{[}{]}{,}{n}{}{,}

\title{Nucleosynthesis in neutrino heated matter: The $\nu p$-process
  and the r-process}

\ShortTitle{Nucleosynthesis in neutrino heated matter}

\author{\speaker{G. Mart{\'i}nez-Pinedo}, A. Keli\'c, K. Langanke,
    K.-H. Schmidt\\
  Gesellschaft f\"ur Schwerionenforschung,\\
  D-64291 Darmstadt, Germany\\
  E-mail: \email{g.martinez@gsi.de}}

\author{D. Mocelj, C. Fr\"ohlich, F.-K. Thielemann, I. Panov,
  T. Rauscher, M. Liebend\"orfer\\
  Department of Physics and Astronomy, University of Basel\\
  Klingelbergstrasse 82, CH-4056 Basel, Switzerland}

\author{N. T. Zinner\\
  Institute for Physics and Astronomy, University of {\AA}rhus,\\
  DK-8000 {\AA}rhus C, Denmark}

\author{B. Pfeiffer\\
  Institute for Nuclear Chemistry, University of Mainz\\
  Fritz-Strassmann-Weg 2, D-55128 Mainz, Germany}

\author{R. Buras and H.-Th. Janka\\
  Max-Planc-Institut f\"ur Astrophysik,
  Karl-Schwarzschild-Strasse 1,\\
  D-85741 Garching, Germany}

\abstract{This manuscript reviews recent progress in our understanding
  of the nucleosynthesis of medium and heavy elements in supernovae.
  Recent hydrodynamical models of core-collapse supernovae show that a
  large amount of proton rich matter is ejected under strong neutrino
  fluxes. This matter constitutes the site of the $\nu p$-process
  where antineutrino absorption reactions catalyze the nucleosynthesis
  of nuclei with $A>64$.  Supernovae are also associated with the
  r-process responsible for the synthesis of the heaviest elements in
  nature. Fission during the r-process can play a major role in
  determining the final abundance patter and in explaining the almost
  universal features seen in metal-poor r-process-rich stars.}

\FullConference{International Symposium on Nuclear Astrophysics - Nuclei in the Cosmos - IX\\
		 25-30 June 2006\\
		 CERN}

\begin{document}

\section{Introduction}\label{sec:intro}

In recent years, observations of metal-poor stars have contributed to
increase our understanding of the nucleosynthesis of medium and heavy
nuclei and its evolution during the history of the galaxy.  Metal-poor
stars with large enhancements of r-process elements (the abundance of
Eu is typically considered to represent the presence of heavy
r-process nuclei) with respect to iron show a variation of two to
three orders of magnitude in the absolute amount of r-process elements
present for stars of similar metallicities~\cite{Cowan.Sneden:2006}.
However the relative abundance of elements heavier that $Z>56$ (but
not including the radioactive actinides) shows a striking consistency
with the observed solar abundances of these
elements~\cite{Cowan.Sneden:2006}.  This consistency does not extend to
elements lighter than $Z=56$ where some variations are observed.  In
most of the cases elements lighter than $Z<56$ are underabundant when
compared with a scaled solar r-process abundance distribution that matches the
observed heavy element abundances~\cite{Cowan.Sneden:2006}.  However,
recent observations of the metal-poor star HD
221170~\cite{Ivans.Simmerer.ea:2006} show that in some cases the
agreement between the scaled solar r-process abundance pattern and the
observed abundances of elements can be extended to elements heavier
than $Z>37$. All these observations indicate that the astrophysical
sites for the synthesis of light and heavy neutron capture elements
are different~\cite{Wasserburg.Busso.Gallino:1996,Qian.Wasserburg:2001}
suggesting two disting r-processes.  Possible sites are supernovae and
neutron-star mergers. The exact site and operation for both types of
r-process is not known, however, there are clear indications that while
the process responsible for the production of heavy elements is
universal~\cite{Wanajo.Ishimaru:2006} the production of lighter
elements (in particular Sr, Y and Zr) has a much more complex Galactic
history~\cite{Travaglio.Gallino.ea:2004}.

Even if the astrophysical site of the r-process(es) is (are) unknown,
it is clear that the process is of primary nature. This means that the
site has to produce both the neutrons and seeds necessary for the
occurrence of a phase with fast neutron captures that characterizes
the r-process~\cite{Cowan.Thielemann.Truran:1991}.  Moreover, in order
to explain the observed abundances of U and Th the neutron-to-seed
ratio needs to be larger than $\sim 100$. Under these conditions
fission of r-process nuclei beyond U and Th can play a mallor role in
explaining the universality of the heavy r-process pattern in
metal-poor stars. This issue will be discussed in
section~\ref{sec:fission}.

In section~\ref{sec:nupprocess} we present a new nucleosynthesis
process that we denote the $\nu p$-process which occurs in proton-rich
matter ejected under explosive conditions and in the presence of
strong neutrino fluxes. This process seems necessary to explain the
observed abundances of light p-nuclei, including $^{92,94}$Mo and
$^{96,98}$Ru.

\section{Nucleosynthesis in proton-rich supernova ejecta}
\label{sec:nupprocess}

Recent hydrodynamical studies of core-collapse supernovae have shown
that the bulk of neutrino-heated ejecta during the early phases (first
second) of the supernova explosion is
proton-rich~\cite{Buras.Rampp.ea:2006,%
  Liebendoerfer.Mezzacappa.ea:2001a,Thompson.Quataert.Burrows:2005}.
Nucleosynthesis studies in this environment have shown that these
ejecta could be responsible for the solar abundances of elements like
$^{45}$Sc, $^{49}$Ti and
$^{64}$Zn~\cite{Froehlich.Hauser.ea:2006,Pruet.Woosley.ea:2005}. Once
reactions involving alpha particles freeze out, the composition in
these ejecta is mainly given by $N=Z$ alpha nuclei and free protons.
Proton captures on this nuclei cannot proceed beyond $^{64}$Ge due to
the low proton separation energy of $^{65}$As and the fact that the
beta-decay half-life of $^{64}$Ge (64~s) is much longer than the
typical expansion time scales (a few seconds). However, the proton
densities and temperatures in these ejecta resemble those originally
proposed for the p-process by B$^2$FH~\cite{Burbidge.Burbidge.ea:1957}.
So it is interesting to ask under which conditions the nucleosynthesis
flow can proceed beyond $^{64}$Ge and contribute to the production of
light p-nuclei like $^{92,94}$Mo and $^{96,98}$Ru that are
systematically underproduced in other
scenarios~\cite{Arnould.Goriely:2003}.

Two recent
studies~\cite{Froehlich.Martinez-Pinedo.ea:2006,Pruet.Hoffman.ea:2006}
have shown that the inclusion of neutrino interactions during the
nucleosynthesis permits a new chain of nuclear reactions denoted $\nu
p$-process in ref.~\cite{Froehlich.Martinez-Pinedo.ea:2006}. In this
process nuclei form at a typical distance of $\sim$ 1000 km from
proto-neutron star where antineutrino absorption reactions proceed on
a time scale of seconds that is much shorter than the typical beta
decay half-lives of the most abundant nuclei present (eg. $^{56}$Ni
and $^{64}$Ge). As protons are more abundant than heavy nuclei,
antineutrino capture occurs predominantly on protons via $\bar{\nu}_e
+ p \rightarrow n + e^+$, causing a residual density of free neutrons
of $10^{14}$--$10^{15}$~cm$^{-3}$ for several seconds, when the
temperatures are in the range 1--3~GK\@.  These neutrons can easily be
captured by neutron-deficient $N\sim Z$ nuclei (for example
$^{64}$Ge), which have large neutron capture cross sections. The
amount of nuclei with $A>64$ produced is then directly proportional to
the number of antineutrinos captured.  While proton capture,
$(p,\gamma)$, on $^{64}$Ge takes too long, the $(n,p)$ reaction
dominates (with a lifetime of 0.25~s at a temperature of 2~GK),
permitting the matter flow to continue to heavier nuclei than
$^{64}$Ge via subsequent proton captures and beta decays till the next
alpha nucleus, $^{68}$Se. Here again $(n,p)$ reactions followed by
proton captures and beta decays permit the flow to reach heavier alpha
nuclei. This process can continue till proton capture reactions freeze
out at temperatures around 1~GK\@. The $\nu p$-process is different to
r-process nucleosynthesis in environtments with $Y_e < 0.5$, i.e.\
neutron-rich ejecta, where neutrino captures on neutrons provide
protons that interact mainly with the existing neutrons, producing
alpha-particles and light nuclei. Proton capture by heavy nuclei is
suppressed because of the large Coulomb
barriers~\cite{Fuller.Meyer:1995,Meyer.Mclaughlin.Fuller:1998}.
Consequently, in r-process environments an enhanced formation of the
heaviest nuclei does not take place when neutrino are present. In
proton-rich ejecta, in contrast to
expectation~\cite{Pruet.Woosley.ea:2005}, antineutrino absorption
produces neutrons that do not suffer from Coulomb barriers and are
captured preferentially by heavy neutron-deficient nuclei.

\begin{figure}
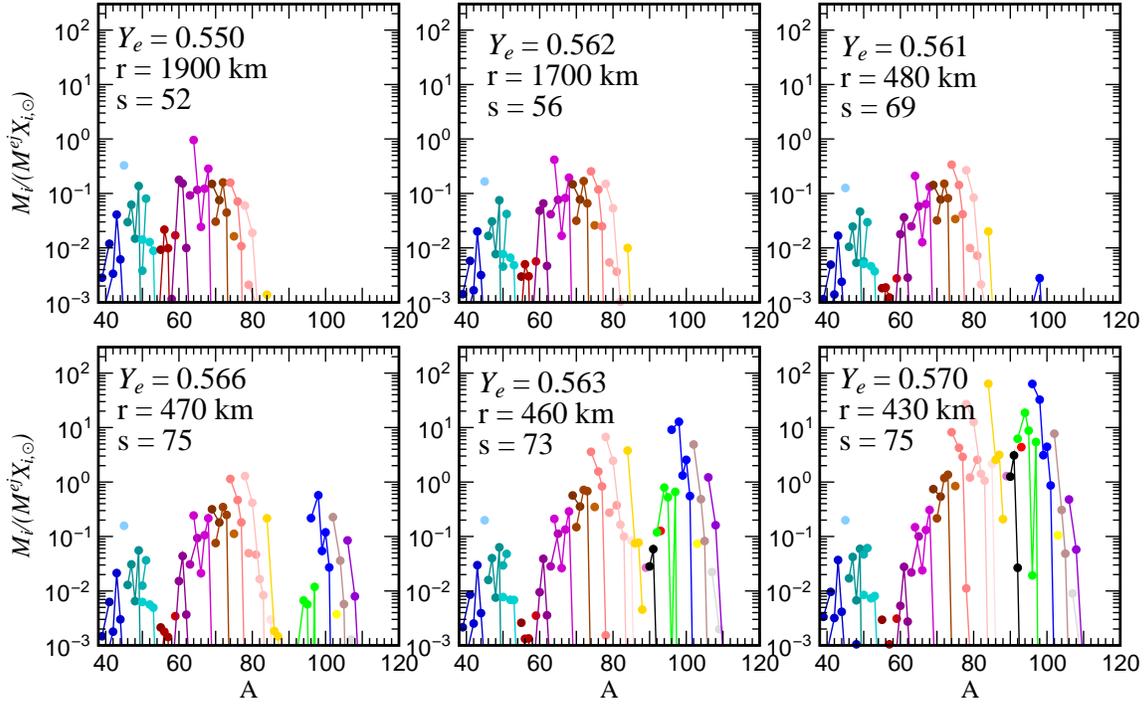

  \ifpdf
  \includegraphics[width=\textwidth]{abund-janka.pdf}
  \else
  \includegraphics[width=\textwidth]{abund-janka.eps}
  \fi
  \caption{Production factors for six hydrodynamical trajectories
    corresponding to the early proton rich wind obtained in the
    explosion of a 15~M$_\odot$
    star~\protect~\cite{Buras.Rampp.ea:2006}. In each panel the radius,
    entropy and $Y_e$ values of matter when the temperature reaches 3
    GK are shown.\label{fig:janka}}
\end{figure}

As discussed above the $\nu p$-process acts in the temperature range
of 1--3 GK. The amount of heavy nuclei synthesized depends on the
ratio of neutrons produced via antineutrino capture to the abundance
of heavy nuclei (this is similar to the neutron-to-seed ratio in the
r-process, see also discussion in~\cite{Pruet.Hoffman.ea:2006}). This
ratio is sensitive to the antineutrino fluence and to the proton to
seed ratio. The first depends mainly on the expansion time scale of
matter and its hydrodynamical evolution. The second is very sensitive
to the proton richness of the material and its entropy.
Figure~\ref{fig:janka} shows the nucleosynthesis resulting from
several trajectories corresponding to the early proton-rich wind from
the protoneutron star resulting of the explosion of a 15~M$_\odot$
star~\cite{Buras.Rampp.ea:2006}. (These trajectories have also been
studied in reference~\cite{Pruet.Hoffman.ea:2006}.) No production of
nuclei above $A=64$ is obtained if antineutrino absorption reactions
are neglected.  Once they are included production of elements above
$A=64$ takes place via the chain of reactions discussed in the
previous paragraph. This allows to extend the nucleosynthesis beyond
Zn producing elements like Ge whose abundance is roughly proportional
to the iron abundance at low metallicities~\cite{Cowan.Sneden:2006}.
The production of light p-nuclei like $^{84}$Sr, $^{94}$Mo and
$^{96,98}$Ru is also clearly seen in figure~\ref{fig:janka}. However,
$^{92}$Mo is still underproduced.  This could be due to the limited
knowledge of masses around $^{92}$Pd~\cite{Pruet.Hoffman.ea:2006}. The
current mass systematics~\cite{Audi.Wapstra.Thibault:2003} predict a
rather low proton separation energy for $^{91}$Rh that inhibits the
production of $^{92}$Pd. Future experimental work in this region
should clarify this issue. However, this could be a feature of the
$\nu p$-process.  In this case, it is interesting to notice that
previous studies have shown that $^{92}$Mo can be produced in slightly
neutron-rich ejecta with $Y_e \approx
0.47$--0.49~\cite{Fuller.Meyer:1995,Hoffman.Woosley.ea:1996}. A recent
study~\cite{Wanajo:2006} has shown that a combination of proton-rich
and slightly neutron-rich ejecta produces all light p-nuclei.
Certainly, much work needs to be done in order to understand the
transition from proton-rich to neutron-rich matter in consistent
supernovae simulations and its dependence with stellar mass.

\section{The role of fission in the r-process}
\label{sec:fission}

The r-process is responsible for the synthesis of at least half of the
elements heavier than Fe. It is associated with explosive scenarios
where large neutrons densities are achieved allowing for the series of
neutron captures and beta decays that constitutes the
r-process~\cite{Cowan.Thielemann.Truran:1991,Cowan.Thielemann:2004}.
The r-process requires the knowledge of masses and beta-decays for
thousands of extremely neutron-rich nuclei reaching even the
neutron-drip line. Moreover, in order to synthesize the heavy
long-lived actinides, U and Th, large neutron to seed ratios are
required ($\sim 100$) allowing to reach nuclei that decay by fission.
Fission can be induced by different processes: spontaneous fission,
neutron induced fission, beta-delayed fission and, if the r-process
occurs under strong neutrino fluxes, neutrino-induced fission. The
role of fission in the r-process has been the subject of many studies
in the past (see ref.~\cite{Panov.Kolbe.ea:2005} and references
therein), however, often only a subset of fission-inducing reactions
was considered and a rather simplistic description of fission yields
was used. It should be emphasized that, if fission really plays a role
in determining the final abundances of the r-process, one needs not
only fission rates but equally important are realistic fission yields
as they determine the final abundances. Our goal has been to improve
this situation by putting together a full set of fission rates
including all possible fission reactions listed above. We use the
Thomas-Fermi fission barriers of reference~\cite{Myers.Swiatecki:1999}
which accurately reproduce the isospin dependence of saddle-point
masses~\cite{Kelic.Schmidt:2005}. The neutron-induced fission rates are
from reference~\cite{Panov.Kolbe.ea:2005}.  Beta-delayed fission rates
are determined based on the FRDM beta-decay
rates~\cite{Moeller.Pfeiffer.Kratz:2003} using an approximate strength
distribution for each decay build on the neutron-emission
probabilities\footnote{\href{http://t16web.lanl.gov/Moller/publications/tpnff.dat}{\url{http://t16web.lanl.gov/Moller/publications/tpnff.dat}}}.
The spontaneous fission rates are determined by a regression fit of
experimental data~\cite{Kodama.Takahashi:1975} to the Thomas-Fermi
fission barriers. For each fissioning nucleus the fission yields are
determined using the statistical code
ABLA~\cite{Gaimard.Schmidt:1991,Benlliure.Grewe.ea:1998}. The fission
yields change from nucleus to nucleus and in a given nucleus depend on
the excitation energy at which fission is induced.

\begin{figure}[htb]
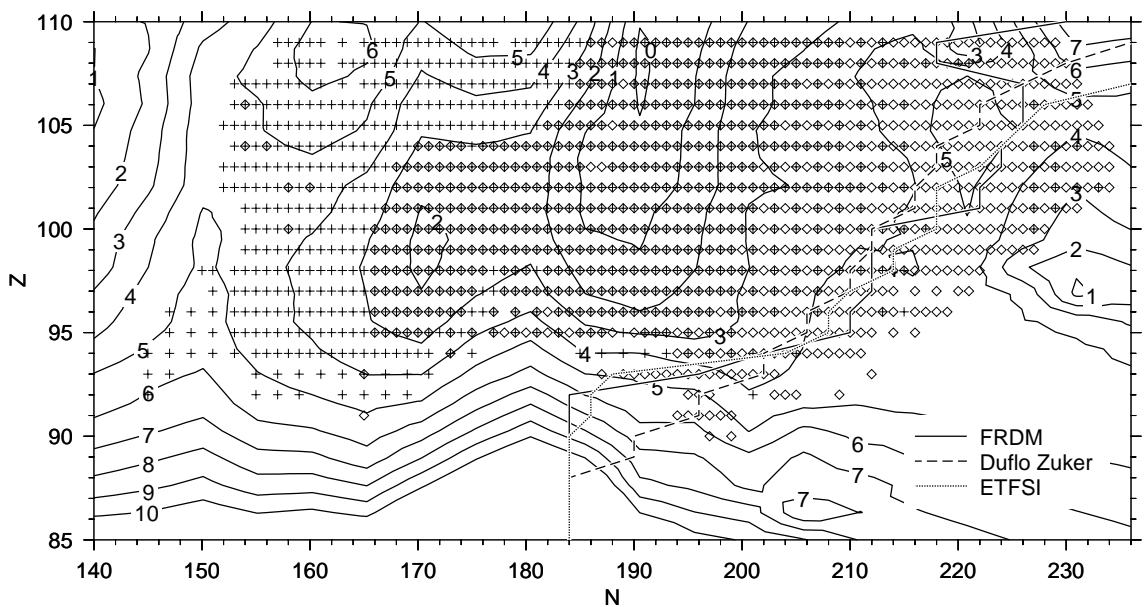

  \ifpdf
  \includegraphics[width=\textwidth]{bf-contour.pdf}
  \else
  \includegraphics[width=\textwidth]{bf-contour.eps}
  \fi
  \caption{Region of the nuclear chart where fission takes place
    during the r-process. The contour lines represent the Thomas-Fermi
    fission barrier heights in MeV. Crosses show the nuclei for which
    neutron-induced fission dominates over $(n,\gamma)$. Diamonds show
    the nuclei for which the spontaneous fission or beta-delayed
    fission operates in a time scale smaller than 1 second. The lines show
    the location for which negative neutron separation energies are
    found in different mass models
    (FRDM\protect~\cite{Moeller.Nix.Kratz:1997},
    ETFSI\protect~\cite{Aboussir.Pearson.ea:1992} and
    Duflo-Zuker\protect~\cite{Duflo.Zuker:1995}).\label{fig:fissionnz}}
\end{figure}

Figure~\ref{fig:fissionnz} shows the region where fission takes place
during the r-process. When the r-process reaches nuclei with $Z\sim
85$--90 matter accumulates at the magic neutron number $N=184$ that
plays a similar role as the standard waiting points at $N=82$ and 126.
Nuclei in this mass range have large fission barriers so that fission
is only possible once matter moves beyond $N=184$. The amount of
matter that is able to proceed beyond this point depends of the
magnitude of the $N=184$ shell gap. The
Duflo-Zuker~\cite{Duflo.Zuker:1995} mass model shows the weakest shell
gap, while masses based on the ETFSI
model~\cite{Aboussir.Pearson.ea:1992} show the stronger shell gap; the
FRDM model~\cite{Moeller.Nix.Kratz:1997} is somewhat in between. Once
matter has passed $N=184$, neutron-induced fission takes place in the
region $Z\sim 90$--95 and $N\sim 190$. Once fission occurs, the main
consequence is that neutrons are mainly captured by fissioning nuclei
that have a larger net capture rate (difference between the capture
and its inverse process). Once a neutron induces a fission the
fissioning nucleus emits around 2--4 neutrons during the fission
process. But a larger amount of neutrons is produced by the decay of
the fission products which have a $Z/A$ ratio similar to the
fissioning nucleus so that they are located closer to the neutron-drip
line than the r-process path. Thus, the fission products will decay
either by photodissociation, $(\gamma,n)$, or beta decays (mainly by
beta-delayed neutron emission) to the r-process path, emitting of
order 8 neutrons per fragment. This implies that each neutron-induced
fission event produces around 20 neutrons.

Once neutrons are exhausted the matter accumulated at $N=184$
beta-decays producing neutrons by beta-delayed neutron emission. These
neutrons induce new fissions in the region $Z\sim 95$, $N\sim 175$
that is fed by beta-decays and produce more neutrons self-enhancing
the neutron-induced fission rate by a mechanism similar to a chain
reaction. The net result is that neutron-induced fission dominates
over beta-delayed or spontaneous fission as it can operate in
time scales of less than a ms for neutron densities above
$10^{18}$~cm$^{-3}$.

The above qualitative arguments which show the role of fission during
the r-process are independent of the fission barriers used. (See for
example figure~2 of reference~\cite{Goriely.Demetriou.ea:2005} where a
figure similar to our figure~\ref{fig:fissionnz} is shown based in the
ETFSI fission barriers~\cite{Mamdouh.Pearson.ea:2001}.) To get a more
quantitative understanding, we have carried out fully dynamical
calculations that resemble the conditions expected in the high-entropy
bubble resulting in a core-collapse supernova explosion. Early
calculations~\cite{Witti.Janka.Takahashi:1994b} failed to produce the
large entropies required for a successful
r-process~\cite{Hoffman.Woosley.Qian:1997}. However, recent
calculations indicate that the high entropies required by the
r-process can be attained~\cite{Burrows.Livne.ea:2006} (see also the
contributions of A. Burrows and A. Arcones). In our calculations, we
assume an adiabatic expansion of the matter, as described in
reference~\cite{Freiburghaus.Rembges.ea:1999}, but using a realistic
equation of state~\cite{Timmes.Arnett:1999}. We adjust the entropy to
produce large enough neutron-to-seed ratios to study the effect of
fission. We notice that the neutron-to-seed ratio does not only
depends on entropy, but also on neutron-richness and expansion time
scale~\cite{Hoffman.Woosley.Qian:1997}.

\begin{figure}[ht]
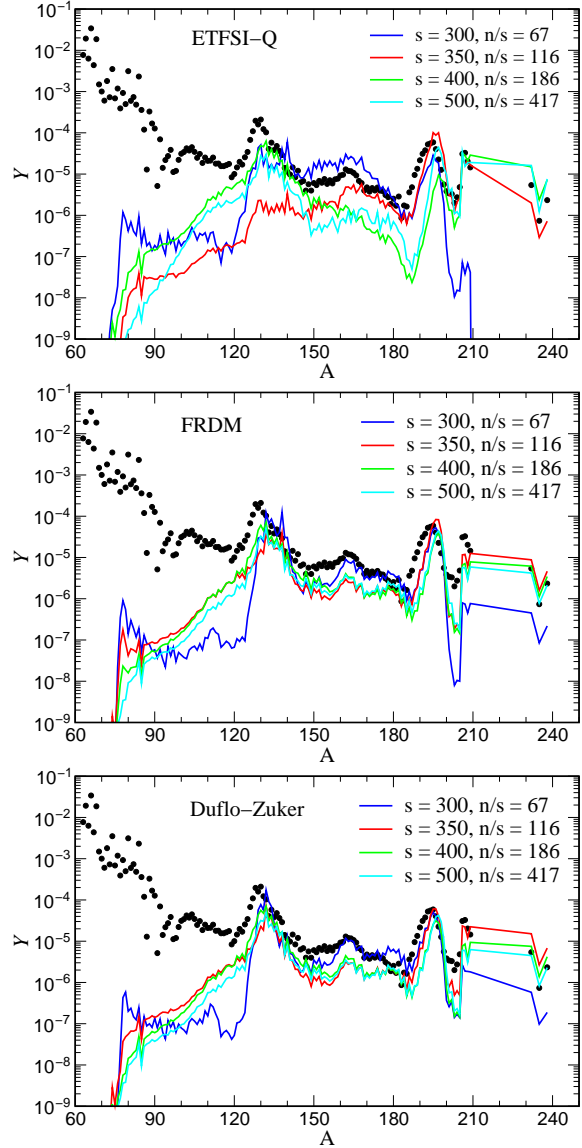

  \centering
  \ifpdf
  \includegraphics[width=0.5\textwidth]{etfsiq-aabund.pdf}\\
  \includegraphics[width=0.5\textwidth]{frdm-aabund.pdf}\\
  \includegraphics[width=0.5\textwidth]{duzu-aabund.pdf}
  \else
  \includegraphics[width=0.5\textwidth]{etfsiq-aabund.eps}\\
  \includegraphics[width=0.5\textwidth]{frdm-aabund.eps}\\
  \includegraphics[width=0.5\textwidth]{duzu-aabund.eps}
  \fi
  \caption{Final r-process abundances obtained in several adiabatic
    expansions using different mass models
    (FRDM\protect~\cite{Moeller.Nix.Kratz:1997},
    ETFSI-Q\protect~\cite{Pearson.Nayak.Goriely:1996} and
    Duflo-Zuker\protect~\cite{Duflo.Zuker:1995}). All the calculation
    are done for a constant expansion velocity of 4500~km
    (corresponding to a dynamical time scale of 50~ms). The product
    $\rho r^3$ is keep constant during the expansion and the
    temperature is determined from the equation of state under the
    condition of constant entropy. The curves are labeled according to
    the entropy and neutron to seed ratio ($n/s$) resulting after the
    alpha-rich freeze-out. The solid circles are a scaled solar
    r-process abundance
    distribution~\cite{Cowan.Pfeiffer.ea:1999}.\label{fig:abund}}
\end{figure}

Figure~\ref{fig:abund} shows the results of our calculations for three
different mass models. While the FRDM and Duflo-Zuker mass models show
a similar trend with increasing neutron-to-seed ratio, the ETFSI-Q
mass model is clearly different. This difference is due to the fact
that the ETFSI-Q mass has a quenched shell gap for $N=82$ and $N=126$,
while the other two mass models show strong shell gaps even close to
the drip line. In the ETFSI-Q mass model the $N=82$ waiting point is
practically absent for the conditions of figure~\ref{fig:abund}. This
allows all matter to pass through $N=82$, incorporating most neutrons
in heavy nuclei and leaving a few free neutrons to induce fission
events. In the other two models, a smaller amount of matter passes the
$N=82$ and $N=126$ waiting points. Once this matter reaches the
fissioning region a large abundance of neutrons is still present that
creates new neutrons by fission allowing the r-process to last for a
longer time and produce a larger fraction of fission fragments. This
explains why the FRDM and Duflo-Zuker mass models produce larger
amounts of matter in the range $A=130$--190, and implies that the
shell structure at $N=82$ is essential for determining the role of
fission in the r-process. Calculations with mass models with strong
shell gaps yield final abundances that are practically independent of
the conditions once the neutron-to-seed ratio is large enough. This
seems to be consistent with metal-poor star observations that show a
universal abundance distribution of elements heavier that
$Z=56$~\cite{Cowan.Sneden:2006}.

Our calculations also show that neutron-induced fission is the major
fission process. For example for the calculations with neutron-to-seed
ratio 186, the percentage of final abundance that has undergone
neutron-induced fission is 36\%, beta-delayed fission 3\% and
neutrino-induced fission 0.3\%. With increasing neutron-to-seed ratio
all the percentages increase but the relative proportions remain
practically constant.

The detection of Th and U in several metal-poor
stars~\cite{Cowan.Sneden:2006} opens the possibility of using the
decay of this elements to estimate the age of the oldest stars in the
galaxy and hence put limits in the age of the Galaxy and the Universe
that are independent of the cosmological model
used~\cite{Cayrel.Hill.ea:2001}. These age estimates need reliable
predictions of the Th and U abundances resulting from the r-process.
Figure~\ref{fig:abund} shows that the absolute amount of Th and U
produced depends of the conditions under which the r-process takes
place. This variations imply that the Th/Eu ratio cannot be used as a
chronometer~\cite{Hill.Plez.ea:2002}.  However, the ratio U/Th is much
less sensitive to the detailed conditions as these nuclei are produced
by alpha decays originating in a similar mass
range~\cite{Schatz.Toenjes.ea:2002,Goriely.Clerbaux:1999}. Using the
mean value of the three calculations with largest entropy shown in
figure~\ref{fig:abund} we obtain a U/Th ratio of $0.59\pm0.02$ for the
FRDM mass model and of $0.56\pm0.04$ for the Duflo-Zuker mass model.
Taking the average of both ratios we determine an age of
$15.4\pm2.4$~Gyr for the metal-poor star CS
31082-001~\cite{Hill.Plez.ea:2002} and of $12.8\pm 6.9$~Gyr for BD
+17$^\circ$3248~\cite{Cowan.Sneden.ea:2002}.

Another interesting issue is the possibility of producing superheavy
elements in the r-process. During our calculations, we produce nuclei
with mass numbers reaching $A=320$ however during the beta decay to
the stability all these nuclei fission. Future work is required to
explore the sensitivity of the potential production of superheavies by
the r-process to different fission barriers. 

\section{Conclusions}
\label{sec:conclusions}

The study of the nucleosynthesis processes responsible for the
production of medium and intermediate elements and their relationship
to supernovae constitutes a challenge to astronomers, astrophysicists
and nuclear physicists. Our current understanding is driven by
high-resolution spectroscopic observations of metal-poor stars that
aim to probe individual nucleosynthesis events. At the same time
progress in the modeling of core-collapse supernovae has improved our
knowledge of explosive nucleosynthesis in supernovae. In particular
the presence of proton-rich ejecta has open the way to find a solution
to the long-standing problem of the origin of light p-nuclei. Further
progress will come from advances in the modeling of the supernovae
explosion mechanism and from improved knowledge of the properties of
the involved nuclei to be studied at future radioactive-ion beam
facilities. These facilities will also open the door to the study many
of the nuclei involved in r-process nucleosynthesis, in particular the
nuclei located near the $N=82$ waiting point that are important in
determining the role of fission in the r-process.  They will also
provide valuable data needed to constrain theoretical models to allow
for more reliable extrapolations to the region of the nuclear chart
where fission takes place during r-process nucleosynthesis.

\bibliographystyle{apsrev}
\bibliography{../biblio/bibliography}

\end{document}